\documentclass[journal]{IEEEtran}

\IEEEoverridecommandlockouts                              

\usepackage{graphicx} 
\usepackage{amsmath} 
\usepackage{amssymb}  
\usepackage{multicol}

\usepackage{subcaption}
\usepackage{color}
\usepackage[usenames,dvipsnames,svgnames,table]{xcolor}

\newtheorem{assumption}{Assumption}
\newtheorem{proposition}{Proposition}
\newtheorem{lemma}{Lemma}
\newtheorem{remark}{Remark}

\usepackage{color}

\title{\LARGE \bf
Integral control of port-Hamiltonian systems: non-passive outputs without coordinate transformation
}

\author{Joel Ferguson, Alejandro Donaire and Richard H. Middleton 
\thanks{Joel Ferguson and Richard H. Middleton are with School of Electrical Engineering and Computer Science and PRC CDSC, The University of Newcastle, Callaghan, NSW 2308, Australia.
        {\tt\small Email: Joel.Ferguson@uon.edu.au, Richard.Middleton@newcastle.edu.au}}%
\thanks{Alejandro Donaire is with the Department of Electrical Engineering and Information Theory and PRISMA Lab, University of Naples Federico II, Napoli 80125, Italy, and the School of Engineering, The University of Newcastle, Callaghan, NSW 2308, Australia. 
{\tt\small Email: Alejandro.Donaire@unina.it}}%
}

\begin{document}

\maketitle
\thispagestyle{empty}
\pagestyle{empty}

\begin{abstract}
In this paper we present a method for the addition of integral action to non-passive outputs of a class of port-Hamiltonian systems. The proposed integral controller is a dynamic extension, constructed from the open loop system, such that the closed loop preserves the port-Hamiltonian form. It is shown that the controller is able to reject the effects of both matched and unmatched disturbances, preserving the regulation of the non-passive outputs. Previous solutions to this problem have relied on a change of coordinates whereas the presented solution is developed using the original state vector and, therefore, retains its physical interpretation. In addition, the resulting closed loop dynamics have a natural interpretation as a Control by Interconnection scheme.
\end{abstract}
\section{INTRODUCTION}
\label{sec1}

Port-Hamiltonian (pH) models describe system dynamics in terms of the energy, interconnection and dissipation structures \cite{Schaft2014}. The physical information readily available from models written in pH form has inspired successful nonlinear control techniques such as energy shaping (ES) and interconnection and damping assignment (IDA) \cite{Ortega2001a,Ortega2004}. Controllers designed with these methods are such that the closed-loop dynamics can be written as a pH system with a desired structure and energy function \cite{Schaft2014}. The closed-loop energy function is chosen to have a minimum at the desired equilibrium point of the control system.

Control by Interconnection (CbI) is a passivity-based control (PBC) methods that considers the dynamics of the controller to be in the pH form \cite{Ortega2008}. The controller is then interconnected to the plant via a power-preserving interconnection which implies that, the closed-loop dynamics are passive if both the plant and controller are passive. Casimir functions (dynamic invariants) are used to collapse the dynamics of the controller and generate a static-feedback law \cite{Ortega2008}. The developments of CbI methods has primarily focused on the stabilisation problem by developing static state-feedback controllers \cite{Schaft2014,Ortega2008,Koopman2012}.

The action of external disturbances on controlled pH systems can produce a shift of the equilibrium or may induce instabilities. Typically, passivity with respect the original input-output pair no longer holds which means that the Hamiltonian cannot be used as a Lyapunov candidate for the disturbed system. However, under the presence of constant disturbances, we can explore the passivity of the so-called incremental model \cite{Pavlov2008,Jayawardhana2007,Jayawardhana2005}. That is, we form a new incremental input about the disturbance and a new incremental output about the corresponding constant output. Then, incremental passivity can be used to obtain stability results of non-zero equilibria.

In the pH framework, the addition of integral action to energy shaping controllers has been proposed to reject the effects of constant disturbances \cite{Donaire2009,Ortega2012}. In the case of passive output regulation, integral action can be applied directly to the output to solve the problem \cite{Ortega2004}. However, often the passive outputs are not the variables of primary interest. An example of this is mechanical system were we are often concerned with the position variables, which are non-passive outputs. A procedure for the addition of integral action to output that are not necessarily passive has been proposed and explored in \cite{Donaire2009,Ortega2012,Romero2013a}. This approach depends on a, possibly implicit, partial coordinate transformation of the states. This transformation may result in a state vector formed by \emph{latent variables} \cite{Willems2007}, and the physical interpretation of the states may be lost, even though the closed loop has a pH form.

This paper presents an alternative approach to design integral action around non-passive outputs for a class of pH systems. The pH structure of the controller is constructed by copying part of the interconnection and dissipation structure of the plant. The controller ensures asymptotic stability of the closed loop to a desired equilibrium and regulation of the non-passive outputs despite the presence of constant disturbances. The interpretation of the closed loop as interconnected systems is compatible with the behavioural approach (see \cite{Willems2007} for a survey on this topic), and the closed loop inherits the passivity properties of the subsystems \cite{Schaft2014}. 
A feature of control designs based on the interconnection of passive systems is that stability of the closed loop holds even under parametric uncertainties, provided that the passive properties remain unchanged \cite{Brogliato2007}.

The remainder of the paper will be structured as follows: Some basic results on port-Hamiltonian systems will be reviewed in Section \ref{sec:PHS}. The main result on the addition of integral action is presented in Section \ref{sec:IA}. The interpretation of the control law as a interconnection of passive systems is discussed in Section \ref{sec:CbI}. The proposed approach for integral control is illustrated using two examples in Section \ref{sec:example}, and conclusions and future work are discussed in Section \ref{sec:concl}.

%

\section{Basic results on port-Hamiltonian systems}\label{sec:PHS}

\subsection{Port-Hamiltonian systems}\label{sec:PHS:ISOPHS}

In this paper we consider port-Hamiltonian systems (more specifically called input-state-output pH systems) which are dynamic systems of the form
\begin{equation}\label{sec:PHS:ISOPHS:model1}
	\begin{split}
		\dot{\mathbf{x}} &= [\mathbf{J}(\mathbf{x})-\mathbf{R}(\mathbf{x})]\nabla\mathcal{H} + \mathbf{G}(\mathbf{x})\mathbf{u} \\
		\mathbf{y} &= \mathbf{G}^\top(\mathbf{x})\nabla\mathcal{H},
	\end{split}
\end{equation}
where $\mathbf{x}\in\mathbb{R}^n$ is the state vector, $\mathbf{J}(\mathbf{x}) = -\mathbf{J}^\top(\mathbf{x})$ is the interconnection matrix, $\mathbf{R}(\mathbf{x}) = \mathbf{R}^\top(\mathbf{x}) \geq 0$ is the dissipation matrix, $\mathcal{H}(\mathbf{x})$ is the Hamiltonian corresponding to the total system energy, $\mathbf{u}\in\mathbb{R}^m$ is the input, $\mathbf{G}(\mathbf{x})$ is the input mapping matrix and $\mathbf{y}\in\mathbb{R}^m$ is the passive output \cite{Schaft2014}.

For most physical systems, the dimension of the input is less than that of the state vector ($m<n$). In this case, the states can be subdivided into those that are within the image of the input mapping matrix and those that are not:
\begin{equation}\label{sec:PHS:ISOPHS:model2}
	\begin{split}
		\begin{bmatrix}
		\dot{\mathbf{x}}_1 \\
		\dot{\mathbf{x}}_2
		\end{bmatrix}
		&= 
		\begin{bmatrix}
		\mathbf{J}_1 - \mathbf{R}_1 & \mathbf{J}_{12}-\mathbf{R}_{12} \\
		-\mathbf{J}_{12}^\top-\mathbf{R}_{12}^\top & \mathbf{J}_{2} - \mathbf{R}_{2}
		\end{bmatrix}
		\begin{bmatrix}
		\nabla_{\mathbf{x}_1}\mathcal{H} \\
		\nabla_{\mathbf{x}_2}\mathcal{H}
		\end{bmatrix}
		+
		\begin{bmatrix}
		\mathbf{G}_1(\mathbf{x}) \\ \mathbf{0}\\
		\end{bmatrix}
		\mathbf{u} \\ 
		\mathbf{y} 
		&= 
		\mathbf{G}_1^\top(\mathbf{x})\nabla_{\mathbf{x}_1}\mathcal{H},
	\end{split}
\end{equation}
where $\mathbf{x}_1,\mathbf{u},\mathbf{y}\in \mathbb{R}^m$, $\mathbf{x}_2\in \mathbb{R}^{n-m}$ and $\mathbf{G}_1(\mathbf{x})$ is full rank.

For systems of the form \eqref{sec:PHS:ISOPHS:model2}, we refer to $\mathbf{y} = \mathbf{G}_1^\top(\mathbf{x})\nabla_{\mathbf{x}_1}\mathcal{H}$ as the passive output of the system due to the natural duality with the input $\mathbf{u}$. It is often the case that the output of the primary interest is $\mathbf{y}_d= \nabla_{\mathbf{x}_2}\mathcal{H}$ or a function of it. As this output is not passive, regulation of the passive output is not sufficient to achieve the desired behaviour of the system in the presence of disturbances. For the remainder of the paper we will refer to $\mathbf{y}_d$ as the non-passive output of the system \eqref{sec:PHS:ISOPHS:model2}.

The class of pH systems \eqref{sec:PHS:ISOPHS:model1} can be extended with the addition of a feedthrough term \cite{Schaft2014}:
\begin{equation}\label{sec:PHS:ISOPHS:model3}
\begin{split}
\dot{\mathbf{x}} &= \left[\mathbf{J}(\mathbf{x})-\mathbf{R}(\mathbf{x})\right]\nabla \mathcal{H}(\mathbf{x}) + \left[\mathbf{G}(\mathbf{x})-\mathbf{P}(\mathbf{x})\right]\mathbf{u} \\
\mathbf{y} &= \left[\mathbf{G}(\mathbf{x})+\mathbf{P}(\mathbf{x})\right]^\top\nabla \mathcal{H}(\mathbf{x}) + \left[\mathbf{M}(\mathbf{x})+\mathbf{S}(\mathbf{x})\right]\mathbf{u},
\end{split}
\end{equation}
subject to
\begin{align}
	\begin{bmatrix}
		\mathbf{R}(\mathbf{x}) & \mathbf{P}(\mathbf{x}) \\
		\mathbf{P}^\top(\mathbf{x}) & \mathbf{S}(\mathbf{x})
	\end{bmatrix}
	&=
	\begin{bmatrix}
		\mathbf{R}(\mathbf{x}) & \mathbf{P}(\mathbf{x}) \\
		\mathbf{P}^\top(\mathbf{x}) & \mathbf{S}(\mathbf{x})
	\end{bmatrix}^\top \geq 0 
	\label{ISOPHS:2}\\
	\begin{bmatrix}
		-\mathbf{J}(\mathbf{x}) & -\mathbf{G}(\mathbf{x}) \\
		\mathbf{G}^\top(\mathbf{x}) & \mathbf{M}(\mathbf{x})
	\end{bmatrix}
	&=
	-
	\begin{bmatrix}
		-\mathbf{J}(\mathbf{x}) & -\mathbf{G}(\mathbf{x}) \\
		\mathbf{G}^\top(\mathbf{x}) & \mathbf{M}(\mathbf{x})
	\end{bmatrix}^\top.  
	\label{ISOPHS:3}
\end{align}
\begin{lemma}\label{ISOPHSwFT}\cite{Schaft2014}
	The pH system \eqref{sec:PHS:ISOPHS:model3} is passive with storage function $\mathcal{H}(\mathbf{x})$, input $\mathbf{u}$ and output $\mathbf{y}$. 
\end{lemma}

%

\subsection{Integral action of passive outputs}\label{PasveIntAction}

Integral action can be applied to passive outputs of pH systems via a dynamic extension \cite{Ortega2004}. To achieve integral action, the controller dynamics are defined as
\begin{equation}\label{IntActionCtrl}
	\begin{split}
		\dot{\boldsymbol{\zeta}} &= \mathbf{K}_I\mathbf{u}_c \\
		\mathbf{y}_c &= \mathbf{K}_I^\top\nabla_{\boldsymbol{\zeta}}\mathcal{H}_c(\boldsymbol{\zeta}),
	\end{split}
\end{equation}
where $\boldsymbol{\zeta} \in \mathbb{R}^{m}$, $\mathbf{K}_I$ is full rank and $\mathcal{H}_c(\boldsymbol{\zeta})$ is a free strictly convex function. The system \eqref{IntActionCtrl} is interconnected with the plant \eqref{sec:PHS:ISOPHS:model2} via the interconnection
\begin{equation}
	\begin{bmatrix}
		\mathbf{u} \\
		\mathbf{u}_c
	\end{bmatrix}
	=
	\begin{bmatrix}
		\mathbf{0} & -\mathbf{G}_1^{-1}(\mathbf{x}) \\
		\mathbf{G}_1^{-\top}(\mathbf{x}) & \mathbf{0}
	\end{bmatrix}
	\begin{bmatrix}
		\mathbf{y} \\
		\mathbf{y}_c
	\end{bmatrix},
\end{equation}
which results in the closed loop dynamics
\begin{equation}
	\begin{split}
		\begin{bmatrix}
		\dot{\mathbf{x}}_1 \\
		\dot{\mathbf{x}}_2 \\
		\dot{\boldsymbol{\zeta}}
		\end{bmatrix}
		&= 
		\begin{bmatrix}
		\mathbf{J}_1 - \mathbf{R}_1 & \mathbf{J}_{12}-\mathbf{R}_{12} & -\mathbf{K}_I^\top \\
		-\mathbf{J}_{12}^\top-\mathbf{R}_{12}^\top & \mathbf{J}_{2} - \mathbf{R}_{2} & \mathbf{0} \\
		\mathbf{K}_I & \mathbf{0} & \mathbf{0}
		\end{bmatrix}
		\begin{bmatrix}
		\nabla_{\mathbf{x}_1}\mathcal{H}_{cl} \\
		\nabla_{\mathbf{x}_2}\mathcal{H}_{cl} \\
		\nabla_{\boldsymbol{\zeta}}\mathcal{H}_{cl}
		\end{bmatrix},
	\end{split}
\end{equation}
where $\mathcal{H}_{cl} \triangleq \mathcal{H} + \mathcal{H}_c$.
The same approach can not be directly extended for the case of non-passive outputs. As there is no control input acting on the $\mathbf{x}_2$ states, it is not possible to interconnect the non-passive output $\mathbf{y}_d=\nabla_{\mathbf{x}_2}\mathcal{H}$ to the controller \eqref{IntActionCtrl} whilst preserving the port-Hamiltonian structure. See \cite{Donaire2009} for a more comprehensive discussion on this limitation.

\subsection{Integral action via change of coordinates}
The addition of integral action to the non-passive outputs \eqref{sec:PHS:ISOPHS:model2} was proposed in \cite{Donaire2009} and further investigated in \cite{Ortega2012}.
Given a plant of the form \eqref{sec:PHS:ISOPHS:model2}, the closed-loop dynamics are proposed to be
\begin{eqnarray}
		\begin{bmatrix}
			\dot{\mathbf{s}} \\
			\dot{\mathbf{x}}_2 \\
			\dot{\boldsymbol{\zeta}}
		\end{bmatrix}
		\hspace{-3mm}&=& \hspace{-3mm}
		\left.
		\begin{bmatrix}
			\mathbf{J}_1 - \mathbf{R}_1 & \mathbf{J}_{12}-\mathbf{R}_{12} & \mathbf{0} \\
			-\mathbf{J}_{12}^\top-\mathbf{R}_{12}^\top & \mathbf{J}_{2} - \mathbf{R}_{2} & -\mathbf{K}_I^\top  \\
			\mathbf{0} & \mathbf{K}_I & \mathbf{0}
		\end{bmatrix}
		\right|_{\mathbf{x}_1=\mathbf{f}(\mathbf{s},\mathbf{x}_2,\boldsymbol{\zeta})} \nonumber\\
		&&\times
		\begin{bmatrix}
			\nabla_{\mathbf{s}}\mathcal{H}_{cl} \\
			\nabla_{\mathbf{x}_2}\mathcal{H}_{cl} \\
			\nabla_{\boldsymbol{\zeta}}\mathcal{H}_{cl}
		\end{bmatrix}, \label{sec:PHS:OldIntAction}
\end{eqnarray}
where $\mathcal{H}_{cl}(\mathbf{s},\mathbf{x}_2,\boldsymbol{\zeta}) = \mathcal{H}(\mathbf{s},\mathbf{x}_2) + \frac12\boldsymbol{\zeta}^\top\mathbf{K}_I^{-1}\boldsymbol{\zeta}$. The partial coordinate transformation $\mathbf{f}:(\mathbf{s},\mathbf{x}_2,\boldsymbol{\zeta})\to\mathbf{x}_1$ is defined by matching the dynamics of $\mathbf{x}_2$ in \eqref{sec:PHS:ISOPHS:model2} with that of \eqref{sec:PHS:OldIntAction}. Then, the control law is computed using the coordinate transformation $\mathbf{f}$ to match the desired dynamics of $\mathbf{s}$, given in \eqref{sec:PHS:OldIntAction}, to $\mathbf{x}_1$ in \eqref{sec:PHS:ISOPHS:model2}. 

\section{Integral action of non-passive outputs without coordinate transformations}
\label{sec:IA}
\subsection{Problem formulation}
We consider the following class of pH systems:
\begin{equation}\label{IA:PHS_OL}
	\begin{split}
		\Sigma_p:
		\begin{bmatrix}
		\dot{\mathbf{x}}_1 \\
		\dot{\mathbf{x}}_2
		\end{bmatrix}
		&= 
		\begin{bmatrix}
		\mathbf{J}_1 - \mathbf{R}_1 & \mathbf{J}_{12} \\
		-\mathbf{J}_{12}^\top & \mathbf{J}_{2} - \mathbf{R}_{2}
		\end{bmatrix}
		\begin{bmatrix}
		\nabla_{\mathbf{x}_1}\mathcal{H} \\
		\nabla_{\mathbf{x}_2}\mathcal{H}
		\end{bmatrix} \\
		&\phantom{-}+
		\begin{bmatrix}
		\mathbf{I} & \mathbf{0}
		\end{bmatrix}^\top
		\mathbf{u}
		-
		\begin{bmatrix}
		\mathbf{I} & \mathbf{0}
		\end{bmatrix}^\top
		\mathbf{d}_1
		-
		\begin{bmatrix}
		\mathbf{0} & \mathbf{I}
		\end{bmatrix}^\top
		\mathbf{d}_2 \\ 
		\mathbf{y} 
		&=
		\nabla_{\mathbf{x}_1}\mathcal{H},
	\end{split}
\end{equation}
where $\dim{\mathbf{x}_1} = m$, $\dim{\mathbf{x}_2} = p \leq m$, $\mathbf{J}_1 = -\mathbf{J}_1^\top$, $\mathbf{J}_2 = -\mathbf{J}_2^\top$, $\mathbf{R}_1 = \mathbf{R}_1^\top > 0$, $\mathbf{R}_2 = \mathbf{R}_2^\top \geq 0$ and $\mathcal{H}:\mathbb{R}^{m+p}\to\mathbb{R}$ is smooth and has a (strict) minimum at $(\mathbf{x}_1^*,\mathbf{x}_2^*)$. $\mathbf{d}_1 \in \mathbb{R}^m$ is a constant, matched disturbance to the system and $\mathbf{d}_2 \in \mathbb{R}^p$ is a constant, unmatched disturbance.
The smooth dependence of the system matrices on $\mathbf{x}_1$ and $\mathbf{x}_2$ is assumed but omitted.

The control objective is to find a dynamic controller $\mathbf{u}=\mathbf{u}(\mathbf{x}_1,\mathbf{x}_2,\boldsymbol{\zeta})$, where $\boldsymbol{\zeta} \in \mathbb{R}^{p}$ is the state of the controller, that ensures asymptotic stability of an equilibrium $(\bar{\mathbf{x}}_1,\mathbf{x}_2^*,\bar{\boldsymbol{\zeta}})$, for constant vectors $\bar{\mathbf{x}}_1\in\mathbb{R}^m$ and $\bar{\boldsymbol{\zeta}}\in\mathbb{R}^p$. Furthermore, the closed loop dynamics should retain the general pH structure.

Notice that we are interested in preserving the component of the original equilibrium associated to $\mathbf{x}_2$ whilst leaving free the component associated to $\mathbf{x}_1$. 

Our approach requires the following assumptions:

\begin{assumption}\label{IA:PHS_OL:Assumptions:1}
	$\mathbf{J}_{12}$ and $\mathbf{R}_1$ are full rank.
\end{assumption}
\begin{assumption}\label{IA:PHS_OL:Assumptions:2}
	$\nabla_{\mathbf{x}_1\mathbf{x}_2}\mathcal{H} = \mathbf{0}$.
\end{assumption}
\begin{assumption}\label{IA:PHS_OL:Assumptions:4}
	$\mathcal{H}$ is strongly convex.
\end{assumption}

\begin{remark}
	The assumption that $\nabla_{\mathbf{x}_1\mathbf{x}_2}\mathcal{H} = \mathbf{0}$ means that there cannot be cross terms between the $\mathbf{x}_1$ states and $\mathbf{x}_2$ states in the open loop Hamiltonian of \eqref{IA:PHS_OL}. This condition is satisfied, for example, by fully actuated electrical machines that are Blondel-Park transformable \cite{Nicklasson1997}, whereby the dynamics are expressed in a rotating frame.
\end{remark}

As discussed in Section \ref{PasveIntAction}, integral action cannot be directly applied to the non-passive output as the pH form will not be preserved. It will be shown in this section that the aforementioned limitation arises due to the selection of the closed-loop Hamiltonian. In this work, the closed-loop Hamiltonian is selected to be the sum of the open-loop Hamiltonian and an additional term which is strictly convex in the difference between $\mathbf{x}_1$ and the controller state $\boldsymbol{\zeta}$. This means that the energy associated with the controller is coupled to the the open-loop plant energy. This coupling is precisely the property that permits disturbance rejection without a coordinate transformation. The resulting control law is a state feedback control law that requires knowledge of the entire state vector.

\subsection{Integral control}\label{sec:IA:n=mCase}
We propose the state feedback control law:
\begin{equation}\label{IA:Controller}
	\begin{split}
		\mathbf{u}
		&=
		(\mathbf{J}_1-\mathbf{R}_1)\nabla_{\mathbf{x}_1}\mathcal{H}_c(\mathbf{E}^\top\mathbf{x}_1-\boldsymbol{\zeta}) \\
		\dot{\boldsymbol{\zeta}}
		&=
		\mathbf{E}^\top\mathbf{J}_{12}\nabla_{\mathbf{x}_2}\mathcal{H},
	\end{split}
\end{equation}
where $\boldsymbol{\zeta} \in \mathbb{R}^p$, $\mathbf{E}\in\mathbb{R}^{m\times p}$ is a constant, full rank matrix and $\mathcal{H}_c(\cdot)$ is a strictly convex function in $\mathbf{z} \triangleq \mathbf{E}^\top\mathbf{x}_1-\boldsymbol{\zeta}$ such that $\nabla_{\mathbf{z}}\mathcal{H}_c:\mathbb{R}^p \to \mathbb{R}^p$ is invertible.

The control law \eqref{IA:Controller}, applied to system \eqref{IA:PHS_OL}, results in the closed loop dynamics
\begin{equation}\label{IA:CLD}
	\begin{split}
		\begin{bmatrix}
			\dot{\mathbf{x}}_1 \\
			\dot{\mathbf{x}}_2 \\
			\dot{\boldsymbol{\zeta}}
		\end{bmatrix}
		&= 
		\underbrace{
			\begin{bmatrix}
				\mathbf{J}_1-\mathbf{R}_1 & \mathbf{J}_{12} & \mathbf{0} \\
				-\mathbf{J}_{12}^\top & \mathbf{J}_{2} - \mathbf{R}_{2} & -\mathbf{J}_{12}^\top\mathbf{E} \\
				\mathbf{0} & \mathbf{E}^\top\mathbf{J}_{12} & \mathbf{0}
			\end{bmatrix}
		}_\mathbf{F(\mathbf{x})}
		\begin{bmatrix}
			\nabla_{\mathbf{x}_1}\mathcal{H}_{cl} \\
			\nabla_{\mathbf{x}_2}\mathcal{H}_{cl} \\
			\nabla_{\boldsymbol{\zeta}}\mathcal{H}_{cl}
		\end{bmatrix} \\
		&\hphantom{{}---}
		-
		\begin{bmatrix}
			\mathbf{I} & \mathbf{0} & \mathbf{0}
		\end{bmatrix}^\top
		\mathbf{d}_1
		-
		\begin{bmatrix}
			\mathbf{0} & \mathbf{I} & \mathbf{0}
		\end{bmatrix}^\top
		\mathbf{d}_2,
	\end{split}
\end{equation}
where $\mathcal{H}_{cl} \triangleq \mathcal{H} + \mathcal{H}_{c}:\mathbb{R}^{m+2p}\to\mathbb{R}$. The definition of $\mathcal{H}_{cl}$ results in the following relationship between the gradients of the open and closed loop Hamiltonians:
\begin{equation}\label{IA:CLDoriginalCoordinates}
	\begin{bmatrix}
		\nabla_{\mathbf{x}_1}\mathcal{H} \\
		\nabla_{\mathbf{x}_2}\mathcal{H} \\
		\nabla_{\mathbf{z}}\mathcal{H}_c
	\end{bmatrix}
	=
	\begin{bmatrix}
		\nabla_{\mathbf{x}_1}\mathcal{H}_{cl} + \mathbf{E}\nabla_{\boldsymbol{\zeta}}\mathcal{H}_{cl} \\
		\nabla_{\mathbf{x}_2}\mathcal{H}_{cl} \\
		-\nabla_{\boldsymbol{\zeta}}\mathcal{H}_{cl}
	\end{bmatrix}.
\end{equation}

The remainder of this section will be focused on the asymptotic behaviour of the closed loop system \eqref{IA:CLD}.
\subsection{Matched disturbance}\label{Matched}
We first consider the case of a matched disturbance only. That is, $\mathbf{d}_1$ is some unknown constant and $\mathbf{d}_2 = \mathbf{0}$. 
\begin{assumption}\label{IA:matched:AssumptionJ1R1const}
	$\mathbf{J}_1$ and $\mathbf{R}_1$ are constant.
\end{assumption}
\begin{assumption}\label{IA:matched:AssumptionM}
The constant $\mathbf{E}$ utilised in the control law \eqref{IA:Controller} can be chosen such that $\mathbf{E}^\top\mathbf{J}_{12}$ is invertible and
	\begin{equation}
		(\mathbf{J}_{12}^{\top*}\mathbf{E})^{-1}\mathbf{J}_{12}^{\top*}
		=
		(\mathbf{J}_{12}^{\top}\mathbf{E})^{-1}\mathbf{J}_{12}^\top.
	\end{equation}
\end{assumption}
\begin{assumption}\label{IA:matched:AssumptionEquilibrium}
	There exists some unique $\bar{\mathbf{x}}_1$ such that
	\begin{equation}\label{IA:matched:AssumptionEquilibriumeq}
		\nabla_{\mathbf{x}_1}\mathcal{H}^*
		=
		\{\mathbf{I}-\mathbf{E}(\mathbf{J}_{12}^{\top*}\mathbf{E})^{-1}\mathbf{J}_{12}^{\top*}\}(\mathbf{J}_1-\mathbf{R}_1)^{-1}\mathbf{d}_1.
	\end{equation}
\end{assumption}
\begin{remark}
	If $\mathbf{J}_1$ and $\mathbf{R}_1$ are not constant, they can be replaced with some constant matrices $\tilde{\mathbf{J}}_1, \tilde{\mathbf{R}}_1$ by the feedback $\mathbf{u} = -(\mathbf{J}_1-\mathbf{R}_1-\tilde{\mathbf{J}}_1+\tilde{\mathbf{R}}_1)\mathbf{y} + \tilde{\mathbf{u}}$, where $\tilde{\mathbf{u}}$ forms the new input for further control design.
\end{remark}
\begin{remark}
	Assumption \ref{IA:matched:AssumptionM} is satisfied in two important cases. Firstly, if $\mathbf{J}_{12}$ is invertible, the assumption is satisfied for any invertible $\mathbf{E}$. Secondly, if $\mathbf{J}_{12}$ is constant, then any constant $\mathbf{E}$ such that $\mathbf{E}^\top\mathbf{J}_{12}$ is invertible will satisfy the assumption. In particular, the selection $\mathbf{E} = \mathbf{J}_{12}$ suffices to satisfy the assumption.
\end{remark}
\begin{remark}
	Assumption \ref{IA:matched:AssumptionEquilibrium} is satisfied if $\mathbf{J}_{12}$ is invertible. Note that the right hand side of \eqref{IA:matched:AssumptionEquilibriumeq} becomes zero, resulting in the equation $\nabla_{\mathbf{x}_1}\mathcal{H}^* = \mathbf{0}$. This equation is satisfied by the equilibrium of the open loop system, $\bar{\mathbf{x}}_1 = \mathbf{x}_1^*$.
\end{remark}

Under Assumptions \ref{IA:matched:AssumptionJ1R1const} - \ref{IA:matched:AssumptionEquilibrium}, the closed-loop dynamics \eqref{IA:CLD}, subject to a matched disturbance only, exhibit the equilibrium point $(\bar{\mathbf{x}}_1,\mathbf{x}_2^*,\bar{\boldsymbol{\zeta}})$, which corresponds to the gradient
\begin{equation}\label{IA:matched:equilibrium}
	\begin{bmatrix}
		\nabla_{\mathbf{x}_1}\mathcal{H}_{cl}^* \\
		\nabla_{\mathbf{x}_2}\mathcal{H}_{cl}^* \\
		\nabla_{\boldsymbol{\zeta}}\mathcal{H}_{cl}^*
	\end{bmatrix}
	=
	\begin{bmatrix}
		(\mathbf{J}_1-\mathbf{R}_1)^{-1}\mathbf{d}_1 \\
		\mathbf{0} \\
		-(\mathbf{J}_{12}^{\top*}\mathbf{E})^{-1}\mathbf{J}_{12}^{\top*}(\mathbf{J}_1-\mathbf{R}_1)^{-1}\mathbf{d}_1
	\end{bmatrix}.
\end{equation}
Using the identity \eqref{IA:CLDoriginalCoordinates}, the gradient of the closed loop Hamiltonian can be related to the gradients of the open-loop Hamiltonian and the controller Hamiltonian as follows:
\begin{equation}\label{IA:matched:equilibriumOriginalCoords}
	\begin{bmatrix}
		\nabla_{\mathbf{x}_1}\mathcal{H}^* \\
		\nabla_{\mathbf{x}_2}\mathcal{H}^* \\
		\nabla_{\mathbf{z}}\mathcal{H}_{c}^*
	\end{bmatrix}
	=
	\begin{bmatrix}
		\{\mathbf{I}-\mathbf{E}(\mathbf{J}_{12}^{\top*}\mathbf{E})^{-1}\mathbf{J}_{12}^{\top*}\}(\mathbf{J}_1-\mathbf{R}_1)^{-1}\mathbf{d}_1 \\
		\mathbf{0} \\
		(\mathbf{J}_{12}^{\top*}\mathbf{E})^{-1}\mathbf{J}_{12}^{\top*}(\mathbf{J}_1-\mathbf{R}_1)^{-1}\mathbf{d}_1
	\end{bmatrix}.
\end{equation}
As $\nabla_{\mathbf{z}}\mathcal{H}_c$ is invertible, \eqref{IA:matched:equilibriumOriginalCoords} corresponds to a unique equilibrium point. 

Importantly, the controller preserves the equilibrium point $\mathbf{x}_2^*$ under the action of matched disturbances. The stability of this equilibrium point is assessed in the following proposition.
\begin{proposition}\label{IA:matched:Prop}
	Consider system \eqref{IA:PHS_OL} subject to unknown, matched disturbance in closed loop with the controller \eqref{IA:Controller}. Then, under Assumptions \ref{IA:PHS_OL:Assumptions:1}-\ref{IA:matched:AssumptionEquilibrium}, the equilibrium of the closed loop, corresponding to the gradient \eqref{IA:matched:equilibrium}, is globally asymptotically stable.
\end{proposition}

\begin{IEEEproof}
	We propose the Lyapunov candidate\footnote{This Lyapunov candidate has been previously utilised for similar analysis in \cite{Alonso2001}, \cite{Jayawardhana2007}.}
	\begin{equation}\label{IA:matched:LyapunovFunction}
		\mathcal{W} = \mathcal{H}_{cl}(\mathbf{w}) - \left[\mathbf{w} - \mathbf{w}^*\right]^\top\nabla\mathcal{H}_{cl}(\mathbf{w}^*) - \mathcal{H}_{cl}(\mathbf{w}^*),
	\end{equation}
	where $\mathbf{w}=\text{col}(\mathbf{x}_1,\mathbf{x}_2,\boldsymbol{\zeta})$. 
	The derivative of $\mathcal{W}$ along the solution of the closed-loop system \eqref{IA:CLD} can be computed as follows:
	\begin{subequations}\label{IA:matched:LyapunovFunctionProof}
		\begin{align}
			\dot{\mathcal{W}}
			&=
			\left\lbrace
				\nabla_\mathbf{w}\mathcal{H}_{cl} - \nabla_\mathbf{w}\mathcal{H}_{cl}^*
			\right\rbrace^\top
			\{\dot{\mathbf{w}} - \mathbf{0}\}
			\label{IA:matched:LyapunovFunctionProof:1} \\
			&=
			\left\lbrace
				\nabla_\mathbf{w}\mathcal{H}_{cl} - \nabla_\mathbf{w}\mathcal{H}_{cl}^*
			\right\rbrace^\top
			\left\lbrace
				\left(
				\mathbf{F}\nabla_\mathbf{w}\mathcal{H}_{cl} - \mathbf{d}
				\right) \right. 
				\nonumber \\
				&\phantom{---}\left.
				- 
				\left(
				\mathbf{F}^*\nabla_\mathbf{w}\mathcal{H}_{cl}^* - \mathbf{d}
				\right)
			\right\rbrace
			\label{IA:matched:LyapunovFunctionProof:3} \\
			&=
			\begin{bmatrix}
				\nabla_{\mathbf{w}}\mathcal{H}_{cl} \\
				\nabla_{\mathbf{w}}\mathcal{H}_{cl}^*
			\end{bmatrix}^\top
			\begin{bmatrix}
				\mathbf{F} & -\mathbf{F}^* \\
				-\mathbf{F} & \mathbf{F}^* \\
			\end{bmatrix}
			\begin{bmatrix}
				\nabla_{\mathbf{w}}\mathcal{H}_{cl} \\
				\nabla_{\mathbf{w}}\mathcal{H}_{cl}^*
			\end{bmatrix} 
			\label{IA:matched:LyapunovFunctionProof:5} \\
			&\leq
			\begin{bmatrix}
				\nabla_{\mathbf{x}_1}\mathcal{H}_{cl} \\
				\nabla_{\mathbf{x}_1}\mathcal{H}_{cl}^*
			\end{bmatrix}^\top
			\begin{bmatrix}
				(\mathbf{J}_1-\mathbf{R}_1) & -(\mathbf{J}_1-\mathbf{R}_1) \\
				-(\mathbf{J}_1-\mathbf{R}_1) & (\mathbf{J}_1-\mathbf{R}_1) \\
			\end{bmatrix}
			\begin{bmatrix}
				\nabla_{\mathbf{x}_1}\mathcal{H}_{cl} \\
				\nabla_{\mathbf{x}_1}\mathcal{H}_{cl}^*
			\end{bmatrix} 
			\nonumber \\
			&\phantom{---}
			+
			\nabla_{\mathbf{x}_2}^\top\mathcal{H}_{cl}\mathbf{J}_{12}^{\top*}\mathbf{E}\nabla_{\boldsymbol{\zeta}}\mathcal{H}_{cl}^*
			+
			\nabla_{\mathbf{x}_2}^\top\mathcal{H}_{cl}\mathbf{J}_{12}^{\top*}\nabla_{\mathbf{x}_1}\mathcal{H}_{cl}^*
			\nonumber \\
			&\phantom{---}
			-
			\nabla_{\boldsymbol{\zeta}}^\top\mathcal{H}_{cl}^*\mathbf{E}^\top\mathbf{J}_{12}\nabla_{\mathbf{x}_2}\mathcal{H}_{cl}
			-
			\nabla_{\mathbf{x}_1}^\top\mathcal{H}_{cl}^*\mathbf{J}_{12}\nabla_{\mathbf{x}_2}\mathcal{H}_{cl}
			\label{IA:matched:LyapunovFunctionProof:6} \\
			&=
			\begin{bmatrix}
				\nabla_{\mathbf{x}_1}\mathcal{H}_{cl} \\
				\nabla_{\mathbf{x}_1}\mathcal{H}_{cl}^*
			\end{bmatrix}^\top
			\begin{bmatrix}
				(\mathbf{J}_1-\mathbf{R}_1) & -(\mathbf{J}_1-\mathbf{R}_1) \\
				-(\mathbf{J}_1-\mathbf{R}_1) & (\mathbf{J}_1-\mathbf{R}_1) \\
			\end{bmatrix}
			\begin{bmatrix}
				\nabla_{\mathbf{x}_1}\mathcal{H}_{cl} \\
				\nabla_{\mathbf{x}_1}\mathcal{H}_{cl}^*
			\end{bmatrix} 
			\nonumber \\
			&\phantom{---}
			+
			\nabla_{\mathbf{x}_2}^\top\mathcal{H}_{cl}\mathbf{J}_{12}^{\top*}\mathbf{E}\nabla_{\boldsymbol{\zeta}}\mathcal{H}_{cl}^*
			-
			\nabla_{\mathbf{x}_2}^\top\mathcal{H}_{cl}\mathbf{J}_{12}^{\top*}\mathbf{E}\nabla_{\boldsymbol{\zeta}}\mathcal{H}_{cl}^*
			\nonumber \\
			&\phantom{---}
			+
			\nabla^\top_{\mathbf{x}_2}\mathcal{H}_{cl}\mathbf{J}_{12}^\top\mathbf{E}
			(\mathbf{J}_{12}^{\top*}\mathbf{E})^{-1}\mathbf{J}_{12}^{\top*}\nabla_{\mathbf{x}_1}\mathcal{H}_{cl}^*
			\nonumber \\
			&\phantom{---}
			-
			\nabla^\top_{\mathbf{x}_2}\mathcal{H}_{cl}\mathbf{J}_{12}^{\top}\mathbf{E}(\mathbf{J}_{12}^{\top}\mathbf{E})^{-1}\mathbf{J}_{12}^\top \nabla_{\mathbf{x}_1}\mathcal{H}_{cl}^*
			\label{IA:matched:LyapunovFunctionProof:7} \\
			&=
			\begin{bmatrix}
				\nabla_{\mathbf{x}_1}\mathcal{H}_{cl} \\
				\nabla_{\mathbf{x}_1}\mathcal{H}_{cl}^*
			\end{bmatrix}^\top
			\begin{bmatrix}
				\mathbf{I} \\ -\mathbf{I}
			\end{bmatrix}
			(\mathbf{J}_1-\mathbf{R}_1)
			\begin{bmatrix}
				\mathbf{I} & -\mathbf{I}
			\end{bmatrix}
			\begin{bmatrix}
				\nabla_{\mathbf{x}_1}\mathcal{H}_{cl} \\
				\nabla_{\mathbf{x}_1}\mathcal{H}_{cl}^*
			\end{bmatrix} 
			\nonumber \\
			&\phantom{---}+
			\nabla^\top_{\mathbf{x}_2}\mathcal{H}_{cl}\mathbf{J}_{12}^{\top}\mathbf{E}\{(\mathbf{J}_{12}^{\top*}\mathbf{E})^{-1}\mathbf{J}_{12}^{\top*}
			\nonumber \\
			&\phantom{---------}
			-(\mathbf{J}_{12}^{\top}\mathbf{E})^{-1}\mathbf{J}_{12}^\top\} \nabla_{\mathbf{x}_1}\mathcal{H}_{cl}^*
			\label{IA:matched:LyapunovFunctionProof:8} \\
			&\leq
			0,
		\end{align}
	\end{subequations}
	where $\mathbf{d} =\begin{bmatrix} \mathbf{I} & \mathbf{0} & \mathbf{0} \end{bmatrix}^\top\mathbf{d}_1$ and we have used Assumptions \ref{IA:matched:AssumptionJ1R1const} and \ref{IA:matched:AssumptionM} and equation \eqref{IA:matched:equilibrium} throughout these equations. 
	
	As \eqref{IA:matched:LyapunovFunctionProof} is negative semi-definite along the trajectories of \eqref{IA:CLD} and $\mathcal{W}$ is strictly convex, there exists a compact set $U \subset \mathbb{R}^n$ defined by	
	\begin{equation}\label{IA:matched:LyapunovFunctionProof2}
	U = \left\lbrace \mathbf{x} \in \mathbb{R}^{3m} | \mathcal{W}(\mathbf{x}) \leq \mathcal{W}(\mathbf{w}_0) \right\rbrace,
	\end{equation}
	which is invariant under the dynamics \eqref{IA:CLD}.
		As $\mathbf{R}_1$ is positive definite, LaSalle's invariance principle implies that the trajectories \eqref{IA:CLD} converge to the maximum invariant set contained within
	\begin{equation}\label{IA:matched:LyapunovFunctionProof3}
		\mathcal{S}=\{ \mathbf{w} |  \nabla_{\mathbf{x}_1}\mathcal{H}_{cl} - \nabla_{\mathbf{x}_1}\mathcal{H}_{cl}^* = \mathbf{0} \}.
	\end{equation}
	Evaluating the dynamics \eqref{IA:CLD}, restricted to the set $\mathcal{S}$, reveal that any solution of the system restricted to $\mathcal{S}$ satisfies
	\begin{equation}\label{IA:matched:LyapunovFunctionProof11}
		\dot{\boldsymbol{\zeta}} = \mathbf{E}^\top\dot{\mathbf{x}}_1.
	\end{equation}
	On the set $\mathcal{S}$ it holds that $\nabla_{\mathbf{x}_1}\mathcal{H}_{cl} = \nabla_{\mathbf{x}_1}\mathcal{H}_{cl}^*$, which is constant. Thus any solution restricted to $\mathcal{S}$ must satisfy
	\begin{equation}\label{IA:matched:LyapunovFunctionProof5}
		\begin{split}
		\frac{d}{dt} \left[ \nabla_{\mathbf{x}_1}\mathcal{H}_{cl} \right]
			&=
			\mathbf{0}.
		\end{split}
	\end{equation}
	Expanding this equation results in:
	\begin{subequations}\label{IA:matched:LyapunovFunctionProof7}
		\begin{align}
			\frac{d}{dt}& \left[ \nabla_{\mathbf{x}_1}\mathcal{H}_{cl} \right] = \nabla_{\mathbf{x}_1}^2\mathcal{H}_{cl}\dot{\mathbf{x}}_1 + \nabla_{\mathbf{x}_1\mathbf{x}_2}\mathcal{H}_{cl}\dot{\mathbf{x}}_2 + \nabla_{\mathbf{x}_1\boldsymbol{\zeta}}\mathcal{H}_{cl}\dot{\boldsymbol{\zeta}} \nonumber \\
			&=
			\nabla_{\mathbf{x}_1}^2\mathcal{H}\dot{\mathbf{x}}_1 + \nabla_{\mathbf{x}_1}^2\mathcal{H}_c\dot{\mathbf{x}}_1 + \nabla_{\mathbf{x}_1\mathbf{x}_2}\mathcal{H}\dot{\mathbf{x}}_2 \nonumber \\
			&\phantom{---}
			 + \nabla_{\mathbf{x}_1\mathbf{x}_2}\mathcal{H}_c\dot{\mathbf{x}}_2 + \nabla_{\mathbf{x}_1\boldsymbol{\zeta}}\mathcal{H}\dot{\boldsymbol{\zeta}} + \nabla_{\mathbf{x}_1\boldsymbol{\zeta}}\mathcal{H}_c\dot{\boldsymbol{\zeta}} \label{IA:matched:LyapunovFunctionProof7:3} \\
			&=
			\nabla_{\mathbf{x}_1}^2\mathcal{H}\dot{\mathbf{x}}_1 + \nabla_{\mathbf{x}_1}^2\mathcal{H}_c\dot{\mathbf{x}}_1 + \nabla_{\mathbf{x}_1\mathbf{x}_2}\mathcal{H}\dot{\mathbf{x}}_2 \nonumber \\ 
			&\phantom{---}
			+ \nabla_{\mathbf{x}_1\boldsymbol{\zeta}}\mathcal{H}_c\dot{\boldsymbol{\zeta}} \label{IA:matched:LyapunovFunctionProof7:4} \\
			&=
			\nabla_{\mathbf{x}_1}^2\mathcal{H}\dot{\mathbf{x}}_1 + \mathbf{E}\nabla_{\boldsymbol{\zeta}}^2\mathcal{H}_c\mathbf{E}^\top\mathcal{H}\dot{\mathbf{x}}_1 + \nabla_{\mathbf{x}_1\mathbf{x}_2}\mathcal{H}\dot{\mathbf{x}}_2 \nonumber \\ 
			&\phantom{---}
			- \mathbf{E}\nabla_{\boldsymbol{\zeta}}^2\mathcal{H}_c\mathbf{E}^\top\dot{\mathbf{x}}_1 \label{IA:matched:LyapunovFunctionProof7:5} \\
			&=
			\nabla_{\mathbf{x}_1}^2\mathcal{H}\dot{\mathbf{x}}_1 + \nabla_{\mathbf{x}_1\mathbf{x}_2}\mathcal{H}\dot{\mathbf{x}}_2 \label{IA:matched:LyapunovFunctionProof7:6} \\
			&=
			\nabla_{\mathbf{x}_1}^2\mathcal{H}\dot{\mathbf{x}}_1, 
			\label{IA:matched:LyapunovFunctionProof7:7}
		\end{align}
	\end{subequations}
	where we have used the symmetry of $\mathcal{H}_c(\mathbf{E}^\top\mathbf{x}_1-\boldsymbol{\zeta})$ together with the identity \eqref{IA:matched:LyapunovFunctionProof11} to make the substitution $-\nabla_{\mathbf{x}_1\boldsymbol{\zeta}}\mathcal{H}_c\dot{\boldsymbol{\zeta}} = \nabla_{\mathbf{x}_1}^2\mathcal{H}_c\dot{\mathbf{x}}_1 = \mathbf{E}\nabla_{\boldsymbol{\zeta}}^2\mathcal{H}_c\mathbf{E}^\top\dot{\mathbf{x}}_1$ in \eqref{IA:matched:LyapunovFunctionProof7:5} and used Assumption \ref{IA:PHS_OL:Assumptions:2} in \eqref{IA:matched:LyapunovFunctionProof7:6}.
	
	Rearranging \eqref{IA:matched:LyapunovFunctionProof7:7}, together with \eqref{IA:matched:LyapunovFunctionProof5} results in
	\begin{equation}\label{IA:matched:LyapunovFunctionProof8}
		\dot{\mathbf{x}}_1 = (\nabla_{\mathbf{x}_1}^2\mathcal{H})^{-1}\mathbf{0} = \mathbf{0},
	\end{equation}
	where $\nabla_{\mathbf{x}_1}^2\mathcal{H}$ is invertible due to Assumption \ref{IA:PHS_OL:Assumptions:4}. Equation \eqref{IA:matched:LyapunovFunctionProof8} further implies that $\dot{\boldsymbol{\zeta}} = \mathbf{0}$ by \eqref{IA:matched:LyapunovFunctionProof11}.
	Considering the $\dot{\mathbf{x}}_1$ dynamics restricted to $\mathcal{S}$,
	\begin{equation}\label{IA:matched:LyapunovFunctionProof9}
		\nabla_{\mathbf{x}_2}\mathcal{H}_{cl} = \nabla_{\mathbf{x}_2}\mathcal{H} = \mathbf{0},
	\end{equation}
	due to \eqref{IA:matched:LyapunovFunctionProof3} and \eqref{IA:matched:LyapunovFunctionProof8}.
	
	By \eqref{IA:matched:LyapunovFunctionProof9}, the dynamics restricted to $\mathcal{S}$ satisfy $\nabla_{\mathbf{x}_2}\mathcal{H} = \mathbf{0}$, which is constant. Taking the time derivative of $\nabla_{\mathbf{x}_2}\mathcal{H} $ it follows that
	\begin{equation}\label{IA:matched:LyapunovFunctionProof10}
		\dot{\mathbf{x}}_2 = (\nabla_{\mathbf{x}_2}^2\mathcal{H})^{-1}\nabla_{\mathbf{x}_2}\mathcal{H} = \mathbf{0},
	\end{equation}
	where $\nabla_{\mathbf{x}_2}^2\mathcal{H}$ is invertible due to the Assumption \ref{IA:PHS_OL:Assumptions:4}.

	Substituting \eqref{IA:matched:LyapunovFunctionProof3}, \eqref{IA:matched:LyapunovFunctionProof9} and \eqref{IA:matched:LyapunovFunctionProof10} into \eqref{IA:CLD} reveals that on the set $\mathcal{S}$
	\begin{equation}\label{IA:matched:LyapunovFunctionProof12}
		\begin{split}
			\nabla_{\boldsymbol{\zeta}}\mathcal{H}_{cl} = -\nabla_{\mathbf{z}}\mathcal{H}_{c}^* &= -(\mathbf{J}_{12}^{\top}\mathbf{E})^{-1}\mathbf{J}_{12}^{\top}(\mathbf{J}_1-\mathbf{R}_1)^{-1}\mathbf{d}_1 \\
			&= -(\mathbf{J}_{12}^{\top*}\mathbf{E})^{-1}\mathbf{J}_{12}^{\top*}(\mathbf{J}_1-\mathbf{R}_1)^{-1}\mathbf{d}_1,
		\end{split}
	\end{equation}
	where we have used Assumption 5. Considering \eqref{IA:matched:LyapunovFunctionProof3}, \eqref{IA:matched:LyapunovFunctionProof9} and \eqref{IA:matched:LyapunovFunctionProof12}, we have recovered the equilibrium gradient  \eqref{IA:matched:equilibrium}. Thus, the maximum invariant set in $\mathcal{S}$ is comprised of a singleton satisfying \eqref{IA:matched:equilibrium}, which implies that the system is globally asymptotically stable to this non-zero equilibrium.
\end{IEEEproof}

\begin{remark}\label{pleqm}
	It is advantageous for $\mathbf{J}_{12}$ to be invertible, so that Assumption \ref{IA:matched:AssumptionM} is satisfied for any choice of $\mathbf{E}$.
	If $\dim \mathbf{x}_2 < \dim \mathbf{x}_1$, it is always possible to add a dynamic extension to the system in order to transform the system into a similar one with $\dim \tilde{\mathbf{x}}_2 = \dim \mathbf{x}_1$. The extended system can be obtained via the feedback law
	\begin{equation}
		\begin{split}
			\dot{\boldsymbol{\gamma}} &= -\mathbf{J}_{12}^\perp \nabla_{\mathbf{x}_1}\mathcal{H} \\
			\mathbf{u} &= (\mathbf{J}_{12}^\perp)^\top \boldsymbol{\gamma} + \mathbf{u}',
		\end{split}
	\end{equation}
	where $\mathbf{J}_{12}^\perp$ is the full-rank left annihilator of $\mathbf{J}_{12}$, which satisfies that $\tilde{\mathbf{J}}_{12} = \begin{bmatrix} \mathbf{J}_{12} & (\mathbf{J}_{12}^\perp)^\top \end{bmatrix}$ is an invertible $m\times m$ matrix. The resulting system has the dynamics
	\begin{equation}
		\begin{split}
			\begin{bmatrix}
				\dot{\mathbf{x}}_1 \\
				\dot{\mathbf{x}}_2 \\
				\dot{\boldsymbol{\gamma}}
			\end{bmatrix}
			&= 
			\begin{bmatrix}
				\mathbf{J}_1-\mathbf{R}_1 & \mathbf{J}_{12} & (\mathbf{J}_{12}^\perp)^\top \\
				-\mathbf{J}_{12}^\top & \mathbf{J}_{2} - \mathbf{R}_{2} & \mathbf{0} \\
				- \mathbf{J}_{12}^\perp & \mathbf{0} & \mathbf{0}
			\end{bmatrix}
			\begin{bmatrix}
				\nabla_{\mathbf{x}_1}\mathcal{H}' \\
				\nabla_{\mathbf{x}_2}\mathcal{H}' \\
				\nabla_{\boldsymbol{\gamma}}\mathcal{H}'
			\end{bmatrix} \\
			&\hphantom{{}---}
			-
			\begin{bmatrix}
				\mathbf{I} \\ \mathbf{0} \\ \mathbf{0}
			\end{bmatrix}
			\mathbf{d}_1
			-
			\begin{bmatrix}
				\mathbf{0} \\ \mathbf{I} \\ \mathbf{0}
			\end{bmatrix}
			\mathbf{d}_2
			+
			\begin{bmatrix}
				\mathbf{I} \\ \mathbf{0} \\ \mathbf{0}
			\end{bmatrix}
			\mathbf{u}',
		\end{split}
	\end{equation}
	where $\mathcal{H}' = \mathcal{H} + \frac12\boldsymbol{\gamma}^\top\boldsymbol{\gamma}$ and $\mathbf{u}'$ is the new input for subsequent control design. This system is now of the form \eqref{IA:CLD} with $\tilde{\mathbf{J}}_{12}$ invertible.
\end{remark}

\subsection{Unmatched disturbance}
We now consider the case of an unmatched disturbance only. That is, $\mathbf{d}_1 = \mathbf{0}$ and $\mathbf{d}_2$ is an unknown constant. In this case $\mathbf{R}_1$ and $\mathbf{J}_1$ can be state-dependant.
\begin{assumption}\label{IA:matched:AssumptionJ12const}
	$\mathbf{J}_{12}$ is constant.
\end{assumption}
\begin{assumption}\label{IA:matched:AssumptionDelHSpan}
	There exists some unique $\bar{\mathbf{x}}_1$ such that
	\begin{equation}
		\nabla_{\mathbf{x}_1}\mathcal{H}^*
		=
		-\mathbf{E}(\mathbf{J}_{12}^{\top}\mathbf{E})^{-1}\mathbf{d}_2.
	\end{equation}
\end{assumption}

Under Assumptions \ref{IA:matched:AssumptionJ12const} and \ref{IA:matched:AssumptionDelHSpan}, the dynamics \eqref{IA:CLD} subject to an unmatched disturbance only, has an equilibrium point $(\bar{\mathbf{x}_1},\mathbf{x}_2^*,\bar{\boldsymbol{\zeta}})$ which corresponds to the gradient
\begin{equation}\label{IA:unmatched:equilibrium}
	\begin{bmatrix}
		\nabla_{\mathbf{x}_1}\mathcal{H}_{cl}^* \\
		\nabla_{\mathbf{x}_2}\mathcal{H}_{cl}^* \\
		\nabla_{\boldsymbol{\zeta}}\mathcal{H}_{cl}^*
	\end{bmatrix}
	=
	\begin{bmatrix}
		\mathbf{0} \\
		\mathbf{0} \\
		-(\mathbf{J}_{12}^{\top}\mathbf{E})^{-1}\mathbf{d}_2
	\end{bmatrix}.
\end{equation}
Using the identity \eqref{IA:CLDoriginalCoordinates}, \eqref{IA:unmatched:equilibrium} can be related to the open-loop Hamiltonian and the controller Hamiltonian:
\begin{equation} \label{equil}
	\begin{bmatrix}
		\nabla_{\mathbf{x}_1}\mathcal{H}^* \\
		\nabla_{\mathbf{x}_2}\mathcal{H}^* \\
		\nabla_{\mathbf{z}}\mathcal{H}_{c}^*
	\end{bmatrix}
	=
	\begin{bmatrix}
		-\mathbf{E}(\mathbf{J}_{12}^{\top}\mathbf{E})^{-1}\mathbf{d}_2 \\
		\mathbf{0} \\
		(\mathbf{J}_{12}^{\top}\mathbf{E})^{-1}\mathbf{d}_2
	\end{bmatrix}.
\end{equation}
As $\nabla_{\mathbf{z}}\mathcal{H}_c$ is invertible, \eqref{equil} corresponds to a valid equilibrium point.
It follows from \eqref{equil} that the value of $\mathbf{x}_2$ at equilibrium is $\mathbf{x}_2^*$ as desired.
\begin{proposition}\label{IA:unmatched:Prop}
	Consider system \eqref{IA:PHS_OL} subject to unmatched, constant disturbance in closed loop with the controller \eqref{IA:Controller}. Then, under Assumptions \ref{IA:PHS_OL:Assumptions:1}-\ref{IA:PHS_OL:Assumptions:4}, \ref{IA:matched:AssumptionJ12const} and \ref{IA:matched:AssumptionDelHSpan},  the equilibrium of the closed loop, corresponding to the gradient \eqref{IA:unmatched:equilibrium}, is globally asymptotically stable.
\end{proposition}

\begin{IEEEproof}
	We again make use of the Lyapunov candidate \eqref{IA:matched:LyapunovFunction}.
	Following the same procedure as the proof of Proposition \ref{IA:matched:Prop}, the derivative of $\mathcal{W}$ along the dynamics of the closed-loop system \eqref{IA:CLD} satisfies:
	\begin{subequations}\label{IA:matched:LyapunovFunctionProof1}
		\begin{align}
			\dot{\mathcal{W}}
			&\leq
			\begin{bmatrix}
				\nabla_{\mathbf{x}_1}\mathcal{H}_{cl} \\
				\nabla_{\mathbf{x}_2}\mathcal{H}_{cl} \\
				\nabla_{\boldsymbol{\zeta}}\mathcal{H}_{cl}^*
			\end{bmatrix}^\top
			\begin{bmatrix}
				(\mathbf{J}_1-\mathbf{R}_1) & \mathbf{0} & \mathbf{0} \\
				\mathbf{0} & \mathbf{0} & \mathbf{J}_{12}^{\top}\mathbf{E} \\
				\mathbf{0} & -\mathbf{E}^\top\mathbf{J}_{12} & \mathbf{0} \\
			\end{bmatrix}
			\begin{bmatrix}
				\nabla_{\mathbf{x}_1}\mathcal{H}_{cl} \\
				\nabla_{\mathbf{x}_2}\mathcal{H}_{cl} \\
				\nabla_{\boldsymbol{\zeta}}\mathcal{H}_{cl}^*
			\end{bmatrix}
			\label{IA:matched:LyapunovFunctionProof1:6} \\
			&=
			-\nabla^\top_{\mathbf{x}_1}\mathcal{H}_{cl}\mathbf{R}_1\nabla_{\mathbf{x}_1}\mathcal{H}_{cl}
			\label{IA:matched:LyapunovFunctionProof1:7} \\
			&\leq
			0,
		\end{align}
	\end{subequations}
	where we made use of Assumption \ref{IA:matched:AssumptionJ12const} and \eqref{IA:unmatched:equilibrium} throughout these equations. 
	
	The proof is completed following the same procedure as Proposition \ref{IA:matched:Prop}, utilising invariance and Assumptions \ref{IA:PHS_OL:Assumptions:1}-\ref{IA:PHS_OL:Assumptions:4}.
\end{IEEEproof}

\subsection{Matched and unmatched disturbance}
Now consider the case where both matched and unmatched disturbances are acting on the system. That is, $\mathbf{d}_1$ and $\mathbf{d}_2$ are non-zero unknown constants. In this case, Assumptions \ref{IA:matched:AssumptionJ1R1const} and \ref{IA:matched:AssumptionJ12const} are required. That is, $\mathbf{J}_1, \mathbf{R}_1$ and $\mathbf{J}_{12}$ must be constant.
\begin{assumption}\label{IA:MixedEquilibrium}
	There exists some unique $\bar{\mathbf{x}}_1$ such that
	\begin{equation}
		\begin{split}
			\nabla_{\mathbf{x}_1}\mathcal{H}^*
			&=
			\{\mathbf{I}-\mathbf{E}(\mathbf{J}_{12}^{\top}\mathbf{E})^{-1}\mathbf{J}_{12}^{\top}\}(\mathbf{J}_1-\mathbf{R}_1)^{-1}\mathbf{d}_1 \\
			&\phantom{------------}
			-\mathbf{E}(\mathbf{J}_{12}^{\top}\mathbf{E})^{-1}\mathbf{d}_2.
		\end{split}
	\end{equation}
\end{assumption}
Under Assumptions \ref{IA:matched:AssumptionJ1R1const}, \ref{IA:matched:AssumptionJ12const} and \ref{IA:MixedEquilibrium}, the equilibrium of the dynamics \eqref{IA:CLD} subject to constant disturbances $\mathbf{d}_1$ and $\mathbf{d}_2$ satisfy
\begin{equation}\label{IA:mixed:equilibrium}
	\begin{bmatrix}
		\nabla_{\mathbf{x}_1}\mathcal{H}_{cl}^* \\
		\nabla_{\mathbf{x}_2}\mathcal{H}_{cl}^* \\
		\nabla_{\boldsymbol{\zeta}}\mathcal{H}_{cl}^*
	\end{bmatrix}
	=
	\begin{bmatrix}
		(\mathbf{J}_1-\mathbf{R}_1)^{-1}\mathbf{d}_1 \\
		\mathbf{0} \\
		-(\mathbf{J}_{12}^{\top*}\mathbf{E})^{-1}\{\mathbf{J}_{12}^{\top*}(\mathbf{J}_1-\mathbf{R}_1)^{-1}\mathbf{d}_1+\mathbf{d}_2\}
	\end{bmatrix},
\end{equation}
which can be related to the open loop Hamiltonian using the identity \eqref{IA:CLDoriginalCoordinates}
Notice that as $\nabla_{\mathbf{x}_2}\mathcal{H}_{cl}^* = \nabla_{\mathbf{x}_2}\mathcal{H}^* = \mathbf{0}$, the equilibrium of the variable $\mathbf{x}_2$ is preserved to its desired values $\mathbf{x}_2^*$ despite the disturbances.
\begin{proposition}\label{IA:mixed:Prop}
	Consider system \eqref{IA:PHS_OL} subject to both matched and unmatched constant disturbances in closed loop with the controller \eqref{IA:Controller}. Then, under Assumptions \ref{IA:PHS_OL:Assumptions:1}-\ref{IA:matched:AssumptionJ1R1const}, \ref{IA:matched:AssumptionJ12const} and \ref{IA:MixedEquilibrium},  the equilibrium of the closed loop, corresponding to the gradient \eqref{IA:mixed:equilibrium}, is globally asymptotically stable.
\end{proposition}

\begin{IEEEproof}
	The proof is obtained by combining the proofs for Propositions \ref{IA:matched:Prop} and \ref{IA:unmatched:Prop}.
\end{IEEEproof}

\section{Interpretation of the integral action controller as C{b}I}\label{sec:CbI}
In this section, the closed-loop dynamics \eqref{IA:CLD} are studied under a coordinate transformation. In the new coordinates, the dynamics coincide with the dynamics presented in \cite{Ferguson2015} with $\mathbf{E} = \mathbf{J}_{12}$, giving a CbI interpretation to the closed loop. This interpretation is of particular interest as in the case that $\mathbf{R}_1 = \mathbf{0}$, disturbance rejection of unmatched disturbances from the non-passive outputs can be achieved without knowledge of the states $\mathbf{x}_2$.

Consider the system \eqref{IA:CLD} under the coordinate transformation
\begin{equation}\label{CbI:coordTransReduced}
	\begin{bmatrix}
		\mathbf{x}_1 \\ \mathbf{x}_2 \\ \mathbf{z}
	\end{bmatrix}
	\triangleq
	\begin{bmatrix}
		\mathbf{I} & \mathbf{0} & \mathbf{0} \\
		\mathbf{0} & \mathbf{I} & \mathbf{0} \\
		\mathbf{E}^\top & \mathbf{0} & -\mathbf{I}
	\end{bmatrix}
	\begin{bmatrix}
		\mathbf{x}_1 \\ \mathbf{x}_2 \\ \boldsymbol{\zeta}
	\end{bmatrix},
\end{equation}
which results in the transformed dynamics
\begin{equation}\label{CbI:CLTransformReduced}
	\begin{split}
		\begin{bmatrix}
			\dot{\mathbf{x}}_1 \\
			\dot{\mathbf{x}}_2 \\
			\dot{\mathbf{z}}
		\end{bmatrix}
		&= 
		\begin{bmatrix}
			\mathbf{J}_1-\mathbf{R}_1 & \mathbf{J}_{12} & (\mathbf{J}_1-\mathbf{R}_1)\mathbf{E} \\
			-\mathbf{J}_{12}^\top & \mathbf{J}_{2} - \mathbf{R}_{2} & \mathbf{0} \\
			\mathbf{E}^\top(\mathbf{J}_1-\mathbf{R}_1) & \mathbf{0} & \mathbf{E}^\top(\mathbf{J}_1-\mathbf{R}_1)\mathbf{E}
		\end{bmatrix} \\
		&
		\phantom{---}\times
		\begin{bmatrix}
			\nabla_{\mathbf{x}_1}\mathcal{H} \\
			\nabla_{\mathbf{x}_2}\mathcal{H} \\
			\nabla_{\mathbf{z}}\mathcal{H}_{c}
		\end{bmatrix}
		-
		\begin{bmatrix}
			\mathbf{I} \\ \mathbf{0} \\ \mathbf{E}^\top
		\end{bmatrix}
		\mathbf{d}_1
		-
		\begin{bmatrix}
			\mathbf{0} \\ \mathbf{I} \\ \mathbf{0}
		\end{bmatrix}
		\mathbf{d}_2.
	\end{split}
\end{equation}
Careful inspection reveals that the system \eqref{CbI:CLTransformReduced} can be obtained as the power-preserving interconnection of the controller
\begin{equation}\label{CbI:Controller}
	\begin{split}
		\Sigma_c: \;
		& \dot{\mathbf{z}} = \mathbf{E}^\top(\mathbf{J}_1 - \mathbf{R}_1)\mathbf{E}\nabla_{\mathbf{z}}\mathcal{H}_c + \mathbf{E}^\top(\mathbf{J}_1 - \mathbf{R}_1)\mathbf{u}_c - \mathbf{E}^\top\mathbf{d}_1 \\
		& \mathbf{y}_c = -(\mathbf{J}_1 - \mathbf{R}_1)\mathbf{E}\nabla_{\mathbf{z}}\mathcal{H}_c + \mathbf{R}_1\mathbf{u}_c,
	\end{split}
\end{equation}
with the plant
\begin{equation}\label{CbI:PHS_OL}
	\begin{split}
		\Sigma_p:
		\begin{bmatrix}
		\dot{\mathbf{x}}_1 \\
		\dot{\mathbf{x}}_2
		\end{bmatrix}
		&= 
		\begin{bmatrix}
		\mathbf{J}_1 & \mathbf{J}_{12} \\
		-\mathbf{J}_{12}^\top & \mathbf{J}_{2} - \mathbf{R}_{2}
		\end{bmatrix}
		\begin{bmatrix}
		\nabla_{\mathbf{x}_1}\mathcal{H} \\
		\nabla_{\mathbf{x}_2}\mathcal{H}
		\end{bmatrix} \\
		&\hphantom{-}+
		\begin{bmatrix}
		\mathbf{I} & \mathbf{0}\\
		\end{bmatrix}^\top
		\mathbf{u}
		-
		\begin{bmatrix}
		\mathbf{I} & \mathbf{0}
		\end{bmatrix}^\top
		\mathbf{d}_1
		-
		\begin{bmatrix}
		\mathbf{0} & \mathbf{I}
		\end{bmatrix}^\top
		\mathbf{d}_2 \\ 
		\mathbf{y} 
		&= 
		\nabla_{\mathbf{x}_1}\mathcal{H},
	\end{split}
\end{equation}
via the interconnection
\begin{equation}\label{CbI:Interconnection}
	\Sigma_{i}: \begin{bmatrix}
	\mathbf{u} \\
	\mathbf{u}_c
	\end{bmatrix}
	=
	\begin{bmatrix}
	\mathbf{0} & -\mathbf{I} \\
	\mathbf{I} & \mathbf{0}
	\end{bmatrix}
	\begin{bmatrix}
	\mathbf{y} \\
	\mathbf{y}_c
	\end{bmatrix}.
\end{equation}
The controller subsystem \eqref{CbI:Controller}  is a pH system of the form \eqref{sec:PHS:ISOPHS:model3} with
\begin{multicols}{2}
	\noindent
	\begin{equation*}
		\begin{split}
			\mathbf{J} &= \mathbf{E}^\top\mathbf{J}_1\mathbf{E} \\
			\mathbf{R} &= \mathbf{E}^\top\mathbf{R}_1\mathbf{E} \\
			\mathbf{S} &= \mathbf{R}_1
		\end{split}
	\end{equation*}
	\begin{equation}\label{CbI:ControllerMatricies}
		\begin{split}
			\mathbf{P} &= \mathbf{E}^\top\mathbf{R}_1 \\
			\mathbf{G} &= \mathbf{E}^\top\mathbf{J}_1 \\
			\mathbf{M} &= \mathbf{0},
		\end{split}
	\end{equation}
\end{multicols}
which is acted upon by a disturbance $-\mathbf{d}_1$. It can be verified that the matrices \eqref{CbI:ControllerMatricies} satisfy \eqref{ISOPHS:2} and \eqref{ISOPHS:3}, thus by Lemma \ref{ISOPHSwFT} the controller is a disturbed, port-Hamiltonian system. 

Under this interpretation, the dissipation term $\mathbf{R}_1$ is viewed as a parameter of the controller subsystem $\Sigma_c$, rather than the plant which restricts our ability to implement the controller as a passive interconnection. Furthermore, in the case of matched disturbances, the controller subsystem \eqref{CbI:Controller} requires knowledge of the disturbance. 

In the case that the open-loop plant has $\mathbf{R}_1 = \mathbf{0}$, integral action can be applied for the rejection of unmatched disturbances using the controller,
\begin{equation}\label{CbI:ControllerCbI}
	\begin{split}
		\Sigma_{\tilde{c}}: \;
		& \dot{\mathbf{z}} = \mathbf{J}_{12}^\top(\mathbf{J}_1 - \mathbf{R}_d)\mathbf{J}_{12}\nabla_{\mathbf{z}}\mathcal{H}_c + \mathbf{J}_{12}^\top(\mathbf{J}_1 - \mathbf{R}_d)\mathbf{u}_c \\
		& \mathbf{y}_c = -(\mathbf{J}_1 - \mathbf{R}_d)\mathbf{J}_{12}\nabla_{\mathbf{z}}\mathcal{H}_c + \mathbf{R}_d\mathbf{u}_c,
	\end{split}
\end{equation}
where $\mathbf{R}_d$ is a positive definite matrix to be chosen. The closed-loop dynamics has the form \eqref{CbI:CLTransformReduced} and, thus, the analysis of section \ref{sec:IA} applies to it.
This formulation has the advantage of not requiring knowledge of the states to be regulated, $\mathbf{x}_2$. This controller was studied in \cite{Ferguson2015} as a disturbance rejection controller for unmatched disturbances.

\section{Examples}
\label{sec:example}
In this section, application of the integral action scheme to a simple RCL circuit is presented to demonstrate the interpretation of the integral action scheme as CbI. The integral action controller is also applied to a class of fully actuated mechanical systems. 
An example application of the proposed action controller to a permanent magnet synchronous motor can be found in \cite{Ferguson2015}.

\subsection{Electrical network}\label{sec:example:RCL}
\begin{figure}[htbp]
	\centering
	\includegraphics[width=0.49\textwidth]{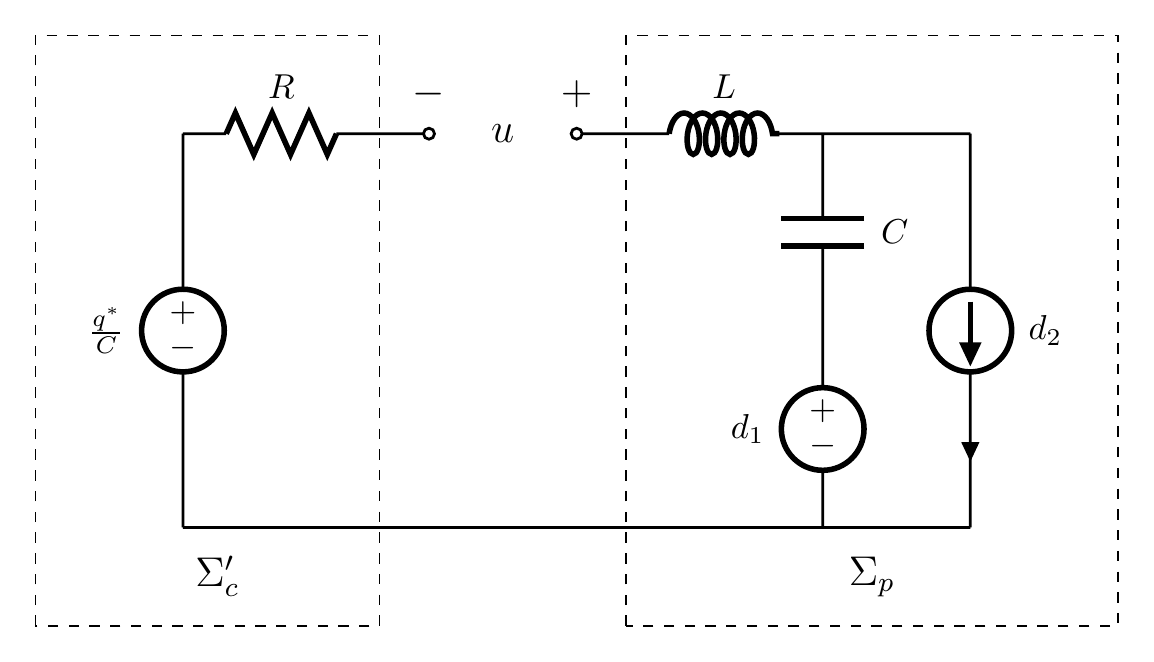}
	\caption{Plant-controller connected through power-preserving interconnection.}
	\label{fig3}
\end{figure}

In this case study, we consider the basic electrical circuit shown in Figure \ref{fig3}. 
This figure represents the feedback interconnection of a plant ($\Sigma_p$) and a controller ($\Sigma_c'$). 
The plant is a LC circuit with a constant source of current acting as an unmatched disturbance and a constant source of voltage acting as a matched disturbance. 
The objective is to regulate the voltage of the capacitor to a desired set-point ($V^* = \frac{q^*}{C}$). 
It can be see that the controller shown in Figure \ref{fig3} achieves the control objective in the case where there are no disturbances and no additional control input, ie. $d_1 = 0$, $d_2 = 0$ and $u=0$.
For a linear inductor, capacitor and resistor, the dynamics of the system are as follows:
\begin{equation}\label{example:8}
	\begin{split}
		\Sigma_{cl}:
		\begin{bmatrix}
			\dot{x}_1 \\
			\dot{x}_2
		\end{bmatrix}
		&=
		\begin{bmatrix}
			-R & -1 \\
			1 & 0
		\end{bmatrix}
		\nabla\mathcal{H}_d(\mathbf{x})
		+
		\begin{bmatrix}
			1 \\
			0
		\end{bmatrix}
		u \\
		&\phantom{---}
		-
		\begin{bmatrix}
			1 \\
			0
		\end{bmatrix}
		d_1 
		-
		\begin{bmatrix}
			0 \\
			1
		\end{bmatrix}
		d_2 \\
		\mathbf{y} 
		&= 
		\begin{bmatrix}
			1 & 0
		\end{bmatrix} 
		\nabla\mathcal{H}_d(\mathbf{x}) \\
		\mathcal{H}_d(\mathbf{x}) 
		&=
		\frac{1}{2L}x_1^2 + \frac{1}{2C}(x_2 - x_2^*)^2,
	\end{split}
\end{equation}
with $\mathbf{x}=\text{col}(x_1,x_2)$ where $x_1$ is the inductor flux and $x_2$ is the capacitor charge.

The integral action controller \eqref{IA:Controller} can be implemented to ensure that $\nabla_{x_2}\mathcal{H}_d = \frac{1}{C}(x_2 - x_2^*)$ converges to zero for any constant disturbances $d_1, d_2$. According the the interpretation of the closed loop provided in Section \ref{sec:CbI}, the resulting dynamics can be interpreted as the interconnection of a plant \eqref{CbI:PHS_OL} with a controller \eqref{CbI:Controller}.

\begin{figure}[htbp]
	\centering
	\includegraphics[width=0.49\textwidth]{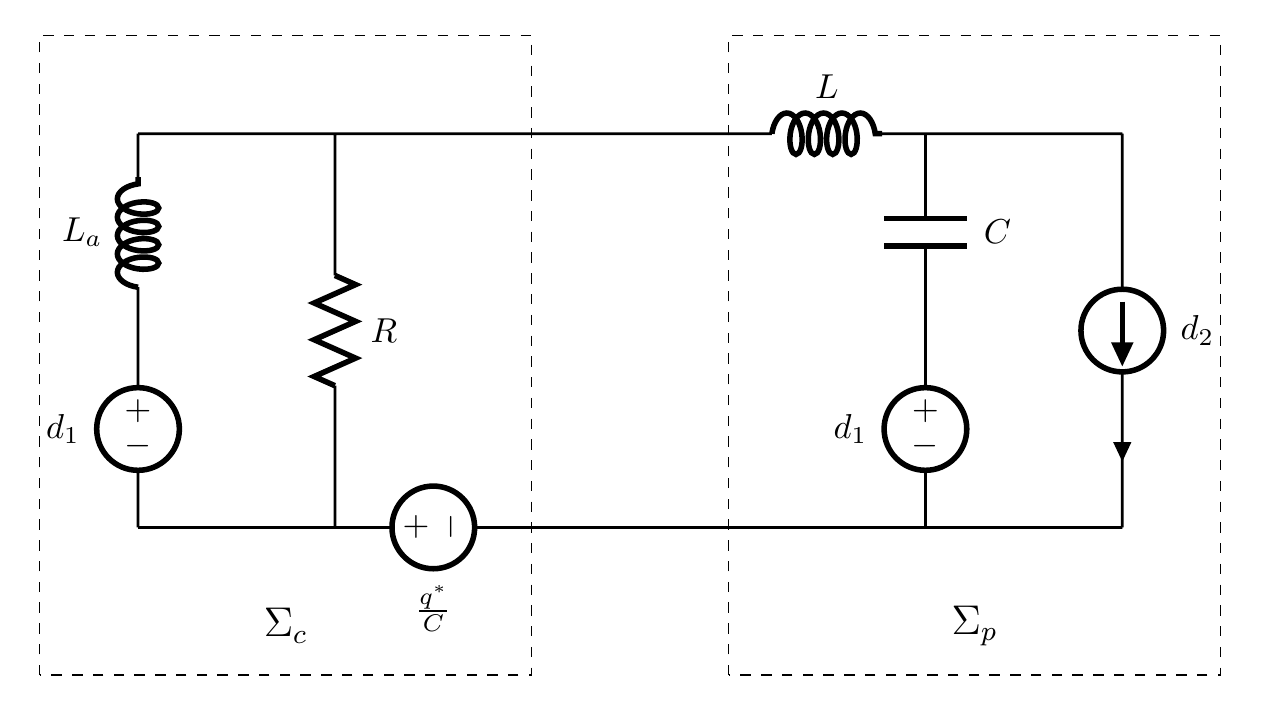}
	\caption{Electrical circuit scheme of a CbI controller interconnected to the plant.}
	\label{fig4}
\end{figure}

For this example, the controller subsystem \eqref{CbI:Controller} has the form
\begin{equation}\label{example:9}
	\begin{split}
		\Sigma_c: \;
		& \dot{\mathbf{z}} = -R\nabla_{\mathbf{z}}\mathcal{H}_c + R\mathbf{u}_c - d_1 \\
		& \mathbf{y}_c = -R\nabla_{\mathbf{z}}\mathcal{H}_c + R\mathbf{u}_c,
	\end{split}
\end{equation}
where $\mathcal{H}_c(\mathbf{z})$ is a free convex function. 
For simplicity, we take $\mathcal{H}_c(\mathbf{z}) = \frac{1}{2L_a}\zeta^2$. 
The integral action controller \eqref{example:9} has the interpretation of a resistor in parallel with a linear inductor, with inductance $L_a$, and voltage source $d_1$ in series. The physical interpretation of the controller $\Sigma_c$ interconnected with the plant $\Sigma_p$ is depicted in Figure \ref{fig4}. 

In the case where there is only an unmatched disturbance acting on the plant, the controller subsystem does not require any knowledge of the disturbances. Thus the controller can be implemented as a CbI scheme. In this electrical example, this is simply the case where $d_1 = 0$. The advantage of implementing the controller in this form is that the controller does not require knowledge of the capacitor voltage, whereas in the form \eqref{IA:Controller}, the controller does require this information.

\subsection{Mechanical systems}\label{sec:example:Mech}
The control law \eqref{IA:Controller} can be applied to a class of mechanical systems to add integral action to the configuration variables. Consider a mechanical system with dynamics
\begin{equation}\label{example:Mechanical:OL}
	\begin{split}
		\begin{bmatrix}
			\dot{\mathbf{p}} \\
			\dot{\mathbf{q}}
		\end{bmatrix}
		&=
		\begin{bmatrix}
			\mathbf{J}(\mathbf{p},\mathbf{q})-\mathbf{D}(\mathbf{p},\mathbf{q}) & -\mathbf{K}^\top(\mathbf{q}) \\
			\mathbf{K}(\mathbf{q}) & \mathbf{0}
		\end{bmatrix}
		\begin{bmatrix}
			\nabla_\mathbf{p}\mathcal{H} \\
			\nabla_\mathbf{q}\mathcal{H}
		\end{bmatrix}
		+
		\begin{bmatrix}
			\mathbf{I} \\
			\mathbf{0}
		\end{bmatrix}
		\mathbf{u} \\
		&\phantom{---}
		-
		\begin{bmatrix}
		\mathbf{I} & \mathbf{0}
		\end{bmatrix}^\top
		\mathbf{d}_1
		-
		\begin{bmatrix}
		\mathbf{0} & \mathbf{I}
		\end{bmatrix}^\top
		\mathbf{d}_2 \\
		\mathbf{y}
		&=
		\frac{\partial\mathcal{H}}{\partial\mathbf{p}} \\
		\mathcal{H}
		&=
		\frac12\mathbf{p}^\top\mathbf{M}^{-1}\mathbf{p}
		+
		\mathcal{V}(\mathbf{q}),
	\end{split}
\end{equation}
where $\mathbf{p}\in\mathbf{R}^m$ is the momenta, $\mathbf{q}\in\mathbf{R}^m$ is the configuration, $\mathbf{J}(\mathbf{p},\mathbf{q}) = -\mathbf{J}^\top(\mathbf{p},\mathbf{q})$ contains the Coriolis and centrifugal terms, $\mathbf{D}(\mathbf{p},\mathbf{q}) = \mathbf{D}^\top(\mathbf{p},\mathbf{q}) > 0$ contains the dissipative terms, $\mathbf{M} = \mathbf{M}^\top > 0$ is the constant mass matrix, $\mathbf{K}(\mathbf{q})$ maps between the reference frame of the momenta and generalised velocities and $\mathcal{V}(\mathbf{q}) > 0$ is the potential energy. $\mathbf{d}_1$ is a vector of force disturbances acting on the system, while $\mathbf{d}_2$ is a vector of velocity disturbances. The mechanical system \eqref{example:Mechanical:OL} is of the form \eqref{IA:PHS_OL} with $\mathbf{x}_1 = \mathbf{p}$ and $\mathbf{x}_2 = \mathbf{q}$.

When controlling mechanical systems, the control objective is typically concerned with the asymptotic behaviour of the configuration variables, $\mathbf{q}$. It is easily verified that Assumptions \ref{IA:PHS_OL:Assumptions:1} - \ref{IA:PHS_OL:Assumptions:4} are satisfied provided that $\mathbf{D} > 0$, $\mathbf{M}$ is constant and $\mathcal{V}(\mathbf{q})$ is strongly convex. As $\dim\mathbf{p} = \dim\mathbf{q}$, Assumption \ref{IA:matched:AssumptionM} is satisfied for any full rank $\mathbf{E}$. Furthermore, as $\nabla_\mathbf{p}\mathcal{H} = \mathbf{M}^{-1}\mathbf{p}$ is invertible, Assumptions \ref{IA:matched:AssumptionEquilibrium}, \ref{IA:matched:AssumptionDelHSpan} and \ref{IA:MixedEquilibrium} are satisfied for each of the relevant disturbance cases. Thus, the control law \eqref{IA:Controller} can be applied to the mechanical system \eqref{example:Mechanical:OL}:
\begin{equation}\label{example:Mechanical:Controller}
	\begin{split}
		\mathbf{u}
		&=
		[\mathbf{J}(\mathbf{p},\mathbf{q})-\mathbf{D}(\mathbf{p},\mathbf{q})]\nabla_{\mathbf{p}}\mathcal{H}_c(\mathbf{E}^\top\mathbf{p}-\boldsymbol{\zeta}) \\
		\dot{\boldsymbol{\zeta}}
		&=
		-\mathbf{K}^\top(\mathbf{q})\nabla_{\mathbf{q}}\mathcal{H}.
	\end{split}
\end{equation}

In the case where $\mathbf{J}(\mathbf{p},\mathbf{q})$ and $\mathbf{D}(\mathbf{p},\mathbf{q})$ are constant, the controller \eqref{example:Mechanical:Controller} can be used to reject the effects of a matched disturbance from the configuration variables. Likewise, in the case where $\mathbf{K}(\mathbf{q})$ is constant, the controller \eqref{example:Mechanical:Controller} can be used to reject the effects of an unmatched disturbance from the configuration variables.


\section{Conclusion}
\label{sec:concl}
This paper presents a method for the addition of integral action that is driven by the non-passive outputs to a pH control systems. Previous methods involve solving (possibly implicit) algebraic expressions to define a suitable coordinate transformation. The control law proposed here is given explicitly, without the need of coordinate transformation. Under suitable assumptions, this control law is able to reject both matched and unmatched disturbances.

The control scheme can be interpreted, under a simple coordinate transformation, as the interconnection of the open-loop plant and a dynamic controller in pH form. This allows the interpretation of the closed loop as a CbI scheme. The controller was applied to a simple RCL circuit and the physical interpretation of the control realised. The control law was also given explicitly for a class mechanical systems.

We notice that Assumption \ref{IA:PHS_OL:Assumptions:2} restricts the applicability of the proposed design to mechanical systems with constant mass matrix. Future research will focus on relaxing assumption 2 and extending the approach in this paper to mechanical systems with non-separable Hamiltonian.


\bibliographystyle{IEEEtran}
\bibliography{IEEEabrv,libraryURLRemoved}

\end{document}